\documentclass[11pt,preprint]{aastex}

\usepackage{graphics,graphicx}
\usepackage{natbib}
\usepackage{float}
\def\apjl{\rm ApJL}
\def\apjs{\rm ApJS}

\def\aap{\rm AAP}

\newcommand{\gtorder}{\mathrel{\raise.3ex\hbox{$>$}\mkern-14mu
            \lower0.6ex\hbox{$\sim$}}}
\newcommand{\ltorder}{\mathrel{\raise.3ex\hbox{$<$}\mkern-14mu
            \lower0.6ex\hbox{$\sim$}}}

\shorttitle{The Case for PSR J1614$-$2230 as a {\it NICER} Target}
\shortauthors{Miller}

\begin{document}

\title{THE CASE FOR PSR J1614$-$2230 AS A {\it NICER} TARGET}

\author{M. Coleman Miller\altaffilmark{1}}

\affil{
{$^1$}{Department of Astronomy and Joint Space-Science Institute, University of Maryland, College Park, MD 20742-2421 USA; miller@astro.umd.edu}
}

\begin{abstract}

The Neutron star Interior Composition ExploreR ({\it NICER}) is expected to launch in early 2017 and will gather X-ray data on neutron stars and other high-energy sources from a berth on the {\it International Space Station}.  Its prime scientific goal is to measure the masses and radii of non-accreting neutron stars via fits to the energy-dependent waveforms produced by the rotation of hot spots on their surfaces.  These measurements will provide valuable input to theoretical models of cold matter beyond nuclear density.  Here we propose that PSR~J1614$-$2230, despite its low count rate, is a promising source to observe with {\it NICER}.  The reason is that {\it XMM-Newton} observations suggest that the fractional oscillation amplitude from PSR~J1614$-$2230 could be high enough that this star cannot be very compact.  We show that if we analyze 0.5~Ms of {\it NICER} data and 0.1~Ms of nearby off-source data and combine that analysis with the known mass of this star, we would find a robust lower limit to the radius with a statistical uncertainty of only $\sim 0.5-0.7$~km.  We also show that even if there is an unmodeled nonthermal component modulated at the pulsation frequency, good statistical fits could rule out significant biases.  The low count rate will make reliable upper limits on the radius difficult, but the lower limit could rule out some equations of state that are currently being discussed.  This analysis would require a good estimate of the non-source background, so {\it Chandra} observations of the vicinity of PSR~J1614$-$2230 would be helpful.

\end{abstract}


\section{INTRODUCTION}
\label{sec:introduction}

The cold matter beyond nuclear density in the cores of neutron stars cannot be replicated in laboratories, and thus there is considerable uncertainty about its properties.  Key guidance about the equation of state of cold dense matter is expected to come from astronomical observations of the masses and especially from the radii of neutron stars.  However, although analyses of X-ray data given particular model assumptions yield circumferential radii on the order of $R\approx 9-14$~km (e.g., \citealt{2013ApJ...762...96B,2013ApJ...772....7G,2013ApJ...765L...5S,2015arXiv150906561N,2016ApJ...820...28O}), the current systematic errors are large; in particular, for many methods the assumptions could be wrong and the inferred radii could be biased by much more than the statistical uncertainty even if the formal statistical fit is good (see, e.g., \citealt{2013arXiv1312.0029M}).  

One method that might be comparatively free of such biases involves fits to the energy-dependent X-ray waveforms of hotter regions (hot spots) on the stellar surface that rotate at close to the rotational frequency of the star.  Existing studies of this method show that even if the actual surface beaming pattern, energy spectrum, temperature distribution over the spot, or spot shape are different from what is assumed in the analysis, there will not be any bias much larger than the statistical uncertainty when the fit is good \citep{2013ApJ...776...19L,2015ApJ...808...31M}.

Waveform fitting will be the focus of the {\it Neutron star Interior Composition ExploreR} ({\it NICER}) mission \citep{2012SPIE.8443E..13G}.  {\it NICER} is expected to launch in early 2017 and will perform deep observations of many sources over its two-year lifetime.  The top targets will be non-accreting neutron stars with X-ray hot spots that are produced by magnetospheric return currents.  These currents are believed to deposit their energy deep enough that the re-radiated energy is thermalized and thus can be described by standard model light-element atmospheres (see \citealt{2013ApJ...762...96B} and references therein), which are likely but not certain to be essentially nonmagnetic for proposed {\it NICER} targets.  The ideal targets will be stars such as PSR~J0437$-$4715 ($\sim 1$ count per second with {\it NICER}), which over a total exposure time of $\sim 10^6$ seconds will produce data of high enough quality to achieve uncertainties of $\sim 5$\% or better on the mass and radius.

Here we suggest that {\it NICER} observations of PSR~J1614$-$2230, despite its low count rate (likely to be $\sim 0.018$ counts per second, from WebPIMMS simulations), have good prospects to place important constraints on the properties of cold dense matter.  {\it XMM-Newton} pn observations of this star \citep{2012A&A...544A.108P} show a $\sim 4\sigma$-significant modulation at the 317.38~Hz rotational frequency.  The fractional root mean squared (frms) amplitude of the modulation is not remarkable by itself, but it is intriguing that of the 1543 counts collected in the $0.4-3$~keV range during the 18.5~ksec exposure, \citet{2012A&A...544A.108P} estimate that only 217 counts came from the star with the rest coming from background; this background was measured using a nearby off-source pointing.  If this is correct, then the X-ray waveform from the star has a very high frms.  As we discuss here, this places an upper limit on the compactness $GM/Rc^2$ of the star, and thus a lower limit on the circumferential radius $R$ given the known gravitational mass $M$ of $1.928\pm 0.017~M_\odot$ (\citealt{2016arXiv160300545F}; for the analysis in this paper we assume $M=1.93~M_\odot$). This is because more compact stars produce greater light deflection, which smooths out the waveform and thus reduces the amplitude.  No similarly rigorous upper limit to the radius can be obtained because the amplitude can also be reduced by other factors (such as a non-pointlike spot, a non-equatorial spot location, or unmodulated emission from the star).  Nonetheless, a lower limit to the radius of a star of this mass would be interesting, particularly because the range of predicted radii at high masses is larger than at low masses (see, e.g., Figure~3 of \citealt{2010Natur.467.1081D}).  We show that a 0.5~Ms {\it NICER} observation of this star combined with 0.1~Ms of off-source observation time to determine the background, would provide data of sufficient quality to limit statistical uncertainties to $\sim 0.5-0.7$~km.

In \S~\ref{sec:methods} we discuss our methods, including our ray-tracing codes and our Bayesian analysis approach.  We also discuss the assumptions we make to maximize the frms for a given compactness; we do this because observations give us a minimum frms from the hot spot, and we want to determine the smallest possible radius that is consistent with that minimum frms.  In \S~\ref{sec:results} we present our results.  We show first that the existing $0.4-3$~keV {\it XMM-Newton} data do not yield interesting constraints if the spot center is on the rotational equator, but that the data disfavor a small radius if the spot center is at the $\sim 40^\circ$ magnetic inclination inferred from {\it Fermi} gamma-ray data \citep{2009ApJ...707..800V}.  Finally, in \S~\ref{sec:summary} we discuss the optimal strategy for the collection and analysis of data from this star.  This strategy includes making {\it Chandra} observations to ascertain whether the non-source background, which dominates the counts, is constant enough over the field that estimates using off-source pointings would not introduce significant systematic errors.

\section{METHODS}
\label{sec:methods}

For PSR~J1614$-$2230, even the {\it NICER} data are unlikely to display the clear harmonic structure in the waveform that is necessary for mass and radius measurements to be accurate to within a few percent.  However, because our focus is on only a lower limit to the radius given the known mass, the frms by itself restricts the radius.  To be as robust as possible in our estimate of the lower limit, we therefore make assumptions about the values of the other parameters of the spot model that maximize the frms.  These assumptions are:

\begin{enumerate}

\item The spot has a very small angular radius.  A larger spot will produce a smoother, lower-amplitude waveform for fixed values of other parameters.  This assumption also has the advantage that for small spots the shape and temperature distribution of the spot does not affect the shape of the waveform \citep{2009ApJ...706..417L,2009ApJ...705L..36L,2015ApJ...811..144B}.  In our calculations, we use a uniform filled circular spot with an angular radius of 0.01~radians.  We also assume that there is only one spot, because multiple spots would decrease the frms at the fundamental oscillation frequency.

\item There are no X-rays from the system that are not modulated at the stellar rotational frequency.  In reality, the non-spot portion of the stellar surface has a nonzero temperature, and in principle there could be X-rays emitted due to, e.g., interaction of the pulsar wind with the companion.  Such X-rays would act as a DC component that would reduce the frms.  Thus we assume that the only DC component will come from unassociated sources, such as background from the {\it International Space Station} or from diffuse astronomical sources.

\item The spot center is very close to the rotational equator, i.e., the spot inclination $\theta_c$ is close to $90^\circ$.  The closer the spot is to the rotational pole, the lower the frms will be (at an extreme, at the pole there will be no rotational modulation).  Gamma-ray analyses suggest that the magnetic inclination is closer to $40^\circ$; if this is the value of $\theta_c$ it would increase significantly the best estimate of the stellar radius (see \S~\ref{sec:results}), but because we want a robust lower limit to $R$ we will usually assume $\theta_c=90^\circ$.  The exception is that when the star is compact enough ($Rc^2/GM\ltorder 3.5$, i.e., $R\ltorder 10$~km for $M=1.93~M_\odot$), lensing will produce caustics when the observer and spot are exactly $180^\circ$ apart that are ruled out by the data.  For such compact stars, we therefore assume $\theta_c=85^\circ$.

\end{enumerate}

Note that a reduction of the observer inclination $\theta_{\rm obs}$ from $90^\circ$ will also reduce the frms.  For PSR~J1614$-$2230 we have an extremely precise measure of the system inclination from measurement of the Shapiro delay: $\theta=89.17^\circ$ \citep{2010Natur.467.1081D}.  Given the standard picture that the star has spun up by accretion, we assume that the stellar rotation axis is aligned with the orbital axis, and thus that $\theta_{\rm obs}= 90^\circ$; if it is not, then the radius is larger than we calculate.

The code we use to produce and analyze synthetic waveforms is the oblate Schwarzschild (OS) code described in \citet{2015ApJ...808...31M}, which evolved from the Schwarzschild+Doppler (S+D) code described in \citet{2013ApJ...776...19L} (note that Appendix~A of \citealt{2013ApJ...776...19L} describes numerous tests of this code).  The OS approximation was introduced by \citet{2007ApJ...654..458C} and \citet{2007ApJ...663.1244M}.  In this approximation, all special relativistic effects (e.g., Doppler shifts and aberration) are included correctly and the oblate shape of the stellar surface is included using a good analytical model, but the exterior spacetime is Schwarzschild rather than including frame-dragging or the effects of mass quadrupoles.  \citet{2007ApJ...663.1244M} show that this produces accurate waveforms for stars that have rotation rates that are $\ltorder 600$~Hz as seen at infinity because although oblateness is second-order and frame-dragging is first-order in rotation, the coefficient for frame-dragging is extremely small at radii comparable to or larger than the stellar radius.  \citet{2007ApJ...663.1244M} and \citet{2015ApJ...808...31M} showed that for $\nu_{\rm rot}\gtorder 300$~Hz the older S+D approximation (in which the star is treated as a sphere but all special relativistic effects are included, see \citealt{1998ApJ...499L..37M,2003MNRAS.343.1301P}; note that
\citealt{1983ApJ...274..846P} and \citealt{1992ApJ...388..138S} discussed
ray-tracing in the Schwarzschild spacetime without including special
relativistic effects) is insufficient, and \citet{2015ApJ...808...31M} also introduced fast codes for the OS approximation, which we use here.  Thus, at the 317.38~Hz rotational frequency of PSR~J1614$-$2230, the OS approximation is both necessary and sufficient.

We combine our code with table lookup of the angle-dependent energy spectrum, which we compute assuming a nonmagnetic pure hydrogen atmosphere and using the public code McPHAC (McGill Planar Hydrogen Atmosphere Code; \citealt{2012ApJ...749...52H}).  The table from which we interpolate has 15 values of $\log$ effective temperature (spaced equally in $\log_{10}(T_{\rm eff}/{\rm K})$ from 5.1 to 6.5), 11 values of $\log$ surface gravity (spaced equally in $\log_{10}(g/{\rm cm~s}^{-2})$ from 13.7 to 14.7), 100 photon energies (spaced equally in $\log_{10}(\hbar\omega/1~{\rm keV})$ from $-1.3$ to 2.0), and 50 angles from the normal (spaced equally from $0.9^\circ$ to $89.1^\circ$).  We verified that interpolation in this table (and extrapolation to $0^\circ$ and $90^\circ$) gives excellent agreement with the results of direct calculation using McPHAC at several intermediate values of $T_{\rm eff}$, $g$, $\hbar\omega$, and the angle from the normal.  A potential issue to study in the future is how close hydrogen atmosphere beaming functions are to helium or carbon beaming functions; the lightest element present should rise and dominate the spectrum, which motivates the focus on hydrogen atmospheres, but because spectral fits to neutron star thermal emission cannot usually distinguish between hydrogen, helium, and carbon (e.g., \citealt{2015A&A...573A..53K}) there is some uncertainty about the composition.

Because of our knowledge of $M$ and $\theta_{\rm obs}$ for this star, and because of our approximations to maximize the frms, our analysis procedure is simplified greatly compared with the analysis needed when no spot or stellar parameters are known a priori (compare \citealt{2015ApJ...808...31M}).  Of the standard parameters in our waveform model ($M$, $R$, $\theta_{\rm obs}$, $\theta_c$, spot angular radius $\Delta\theta$, spot effective temperature $T_{\rm eff}$, and distance $d$ to the star), two are known ($M$ and $\theta_{\rm obs}$), one is assumed ($\theta_c=90^\circ$), and two are degenerate with each other ($\Delta\theta$ and $d$ both only affect the total flux from the spot, given that for our assumed $\Delta\theta\ll 1$ the waveform shape is independent of $\Delta\theta$).  Thus, to construct a posterior probability density for $R$, we need to marginalize over $T_{\rm eff}$ and some combination of $\Delta\theta$ and $d$.  The combination we choose is the total flux from the spot; that is, we pick a fiducial distance and a fiducial spot angular radius and then the model parameter over which we marginalize is the arbitrary factor $f_{\rm mult}$ by which we multiply the spot flux.  

The new aspect of this analysis is that we need to use information about the total background count rate and spectrum.  If, as expected for many {\it NICER} targets, overtones are observed with enough precision to yield a tight constraint on the mass and radius rather than just an upper limit on the compactness, the background mainly contributes statistical fluctuations.  Thus, knowledge of the expected background improves only moderately the measurement precision of $M$ and $R$ when few-percent measurements are possible (see \citealt{2013ApJ...776...19L}, Figures~2(c) and 5(a)).

In contrast, because for PSR~J1614$-$2230 we expect that almost all of our information will come from the frms, in this special case knowledge of the background is essential.  An estimate of the background count rate and spectrum can be obtained using pointings that are angularly close to the source but do not include known sources.  It would therefore be extremely useful to test the spatial uniformity of the background near PSR~J1614$-$2230 using {\it Chandra} observations (see \S~\ref{sec:summary}).  Once we have an estimate of the energy-dependent count rate and uncertainties for the emission {\it not} associated with PSR~J1614$-$2230, we can fold that into our analysis and marginalize over the background.  Note that we model the background rather than simply subtracting it, which would lead to statistically incorrect results.

In our Bayesian analysis, for a given energy channel $i$ the non-source background $B_i$ is treated as a model parameter.  We assume that this background does not vary in a way commensurate with the stellar rotational frequency, and that therefore it is independent of the rotational phase.  If the non-source background is observed for some duration $t_{\rm back}$ and returns a total of $N_i$ counts in channel $i$, then the best estimate of the background $\langle B_i\rangle$ in that channel for a source observation of duration $t_{\rm source}$ is of course $(t_{\rm source}/t_{\rm back})N_i$.  To get the prior probability distribution for $B_i$, we note that the number of counts in each energy channel will be large enough that we can assume Gaussian statistics.  Because the background will be independent of phase, the fractional uncertainty for the background in each phase will be $1/\sqrt{\langle B_i\rangle}$.  That is, even though after folding in phase the expected number of background counts in a given phase bin will be $\langle B_i\rangle/N_{\rm phase}$ if there are $N_{\rm phase}$ phase bins, the uncertainty in the expected number of background counts in that bin will be $\sqrt{\langle B_i\rangle}/N_{\rm phase}$ rather than $\sqrt{\langle B_i\rangle/N_{\rm phase}}$, because the information about the background comes from all phases rather than a specific phase.

Thus, the prior probability distribution of the number of background counts $B_{ij}$ in energy channel $i$ and phase bin $j$ is
\begin{equation}
P(B_{ij})={1\over{\sigma_i\sqrt{2\pi}}}e^{-(B_{ij}-\langle B_i\rangle/N_{\rm phase})^2/2\sigma_i^2}\; ,
\end{equation}
where $\sigma_i=\sqrt{\langle B_i\rangle}/N_{\rm phase}$.  The Poisson likelihood of data $d_{ij}$ given the model and parameter values is
\begin{equation}
{\cal L}=\prod_{i,j}{(S_{ij}+B_{ij})^{d_{ij}}\over{d_{ij}!}}e^{-(S_{ij}+B_{ij})}
\end{equation}
for model source counts $S_{ij}$, and the posterior probability density for $R$ is proportional to
\begin{equation}
Q(R)\propto\int {\cal L}P(R)P(T_{\rm eff})P(f_{\rm mult})P(B_{ij})dT_{\rm eff}df_{\rm mult}dB_{ij}\; .
\end{equation}
Here, $P(R)$ is the prior probability density for $R$ (which for us is flat between $Rc^2/GM=3.1$, or 8.83~km for $M=1.93~M_\odot$, and $Rc^2/GM=5.3$, or 15.1~km for $M=1.93~M_\odot$).  Similarly, $P(T_{\rm eff})$ is the prior probability density for $T_{\rm eff}$, which we take to be flat from $kT_{\rm eff}=0.08$~keV to $kT_{\rm eff}=0.12$~keV for our synthetic data runs in which the real value of $kT_{\rm eff}$ is 0.1~keV (we find that temperatures outside this range give poor fits), and $P(f_{\rm mult})$ is flat in $\ln f_{\rm mult}$ from $-4$ to 10.

\section{RESULTS}
\label{sec:results}

In this section we discuss the specific results of our calculations.  We start with an analysis of the $0.4-3$~keV {\it XMM-Newton} data for PSR~J1614$-$2230 obtained by \citet{2012A&A...544A.108P}, and show that if $\theta_c=90^\circ$ the lower limit on the radius is not constraining.  If instead $\theta_c=40^\circ$ as inferred from gamma-ray data \citep{2009ApJ...707..800V}, then we already have somewhat interesting limits.  Through an analysis of synthetic data, we then explore the limits that could be obtained using {\it NICER} observations.  Finally, we investigate two possible sources of systematic error: unmodeled nonthermal emission that is modulated at the stellar rotational frequency and a systematic under- or over-estimate of the background.  Although a more extensive analysis should be performed, we find it encouraging that the result for both types of errors does not show a significant bias in the inferred lower limit to the radius when the fit is statistically good.

\subsection{Analysis of {\it XMM-Newton} data}
\label{sec:XMM}

\citet{2012A&A...544A.108P} obtained 18.5~ksec of {\it XMM-Newton} pn data on PSR~J1614$-$2230 and because of the small number of counts, they binned all the $0.4-3$~keV data and reported the counts in 16 equally spaced phase bins (see Figure~6 of \citealt{2012A&A...544A.108P}; for the analysis in the current paper, Natalie Webb kindly provided the table of counts per phase bin).  We therefore applied the method of \S~\ref{sec:methods} to this single wide energy bin (weighted by the effective area as a function of photon energy of the {\it XMM-Newton} pn medium filter), and we marginalized over surface comoving effective temperatures $kT_{\rm eff}=0.05$~keV to $kT_{\rm eff}=0.195$~keV as representative values based on the example fits given in Table~3 of \citet{2012A&A...544A.108P}; when fit to a Planck spectrum at infinity, these correspond to observed temperatures of $\sim 0.05$~keV to $\sim 0.27$~keV for our assumed range of $Rc^2/GM=3.1$ to 5.3.  We also assumed an interstellar column of $N_H=2.4\times 10^{21}~{\rm cm}^{-2}$, which is consistent with the fits from \citet{2012A&A...544A.108P}.  Note that the radii found for the spots by \citet{2012A&A...544A.108P} are much smaller than the stellar radius; these are blackbody fits rather than the hydrogen atmosphere fits we use, but if the spot is indeed small then as we discussed earlier there is minimal dependence of the shape of the waveform on the shape or temperature distribution of the spot.

\begin{figure}[!htb]
\begin{center}
\plotone{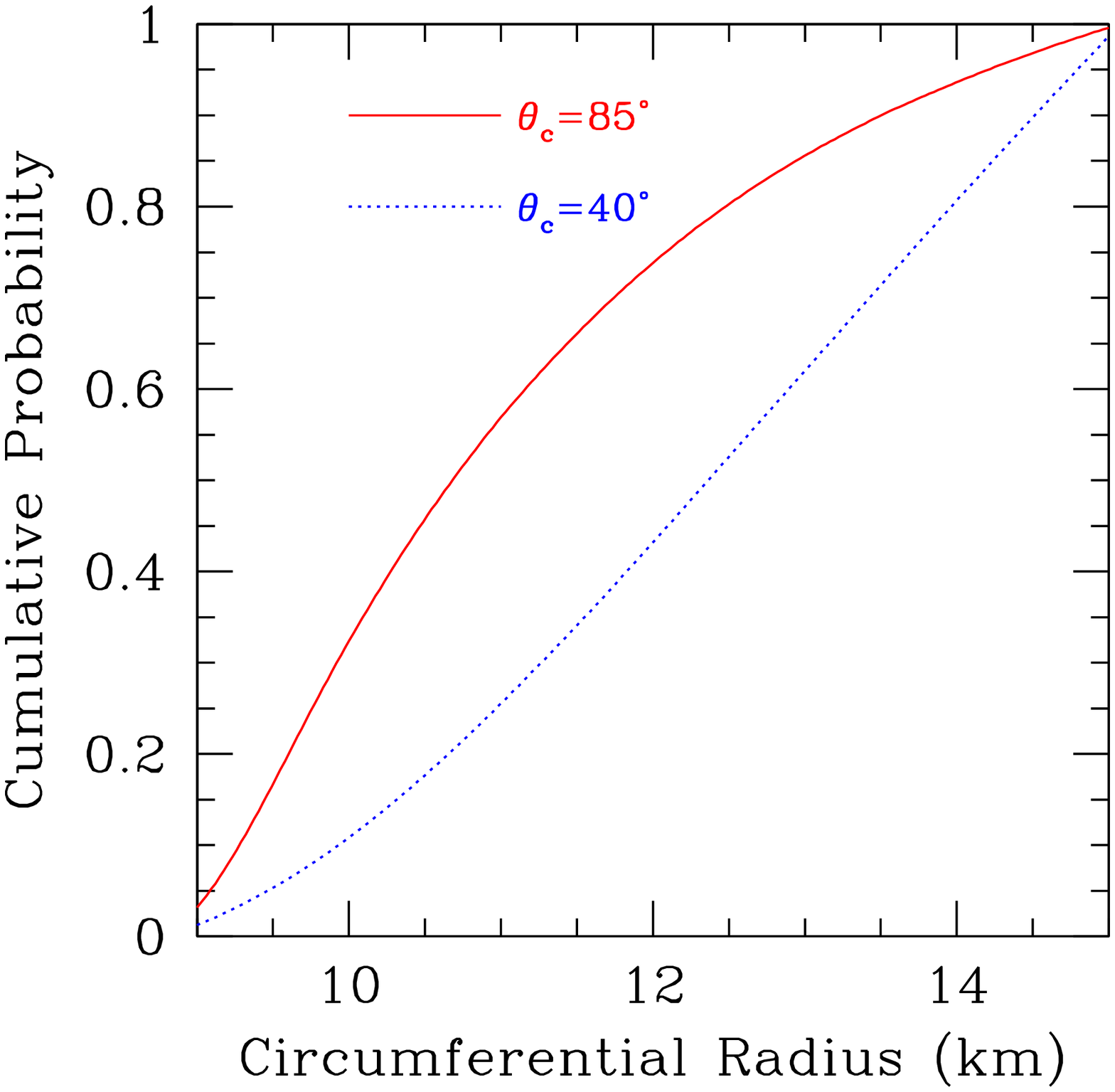}
\vskip-1.0cm
\caption{Cumulative probability of the circumferential radius for a hydrogen atmosphere model of the $0.4-3$~keV {\it XMM-Newton} pn data obtained for PSR~J1614$-$2230 by \citet{2012A&A...544A.108P}.  The solid red line shows the result for a conservative assumption of $\theta_c=85^\circ$ (i.e., the spot is close to the rotational equator; note that for $\theta_c=90^\circ$ and the known $\theta_{\rm obs}\approx 90^\circ$, $Rc^2/GM\ltorder 3.5$, or $R\ltorder 10$~km for the known mass $M=1.93~M_\odot$, produces caustics that are ruled out by the data) and the dotted blue line shows the result for $\theta_c=40^\circ$, which is the magnetic inclination angle found by \citet{2009ApJ...707..800V} from analysis of {\it Fermi} gamma-ray data.  This figure shows that if the spot is closer to the pole than to the equator then the limits are currently somewhat interesting ($R>10$~km with $\sim 10$\% probability), but if the spot is at the equator then no realistic radius is significantly disfavored.  Note that in this figure and the others we present, there are no meaningful upper limits to the radius because for larger radii a smaller $\theta_c$, a larger spot, or an extra unmodulated emission could produce statistically acceptable fits to the real or synthetic data.
}
\label{fig:XMM}
\end{center}
\end{figure}

We show our results in Figure~\ref{fig:XMM}, for the conservative $\theta_c=85^\circ$ (as discussed earlier, $\theta_c=90^\circ$ is strongly ruled out for $Rc^2/GM\ltorder 3.5$ and $\theta_{\rm obs}=90^\circ$ because of the presence of a lensing caustic) and for $\theta_c=40^\circ$, which is the magnetic inclination angle found by \citet{2009ApJ...707..800V} using fits to {\it Fermi} gamma-ray data (but note that from Figures~18(g) and (h) of \citealt{2009ApJ...707..800V} there are substantial residuals in the fit).  The lower limits to the radius are not tight for $\theta_c=85^\circ$, but if $\theta_c=40^\circ$ then they become moderately interesting ($R>10$~km at $\sim 90$\% confidence).  Note that the statistical quality of the fits is the same for $\theta_c=40^\circ$ as it is for $\theta_c=85^\circ$.  This is one reason that for this source we can only get lower limits to the radius; for larger radii, decreasing $\theta_c$, increasing the spot size, or adding unmodulated emission from the star would all result in statistically acceptable fits.

\subsection{Analysis of synthetic {\it NICER} data}

For our synthetic {\it NICER} data sets, we assume a 0.5~Ms observation of PSR~J1614$-$2230 combined with a 0.1~Ms observation of nearby blank fields to estimate the background.  We focus on the $0.4-3$~keV energy range.  We use energy channels of width 0.1~keV, which is the expected resolution of {\it NICER}.  We use a table of effective area versus photon energy that was kindly provided by Zaven Arzoumanian.  At the expected $\sim 0.018$~c/s count rate for PSR~J1614$-$2230 this implies that we would detect $\sim$9000 counts from the source, and at the expected 0.2~c/s background count rate (Z. Arzoumanian 2016, personal communication; note that some of this comes from gamma-ray sources aboard the {\it International Space Station} itself) we would expect roughly 100,000 background counts during the source observation, and 20,000 background counts during the observation of blank fields.  We make the illustrative assumption that the background has a photon count spectrum $dN/dE\propto E^{-2}$, and we generate the blank-field background synthetic data independently from the on-source background synthetic data.  For this section, we assume that the background does not vary in a way that is commensurate with the stellar rotational frequency, which means that the expected background rate should be independent of rotational phase.  Other aspects of the simulation are described in \S~\ref{sec:methods}.

\begin{figure}[!htb]
\begin{center}
\plotone{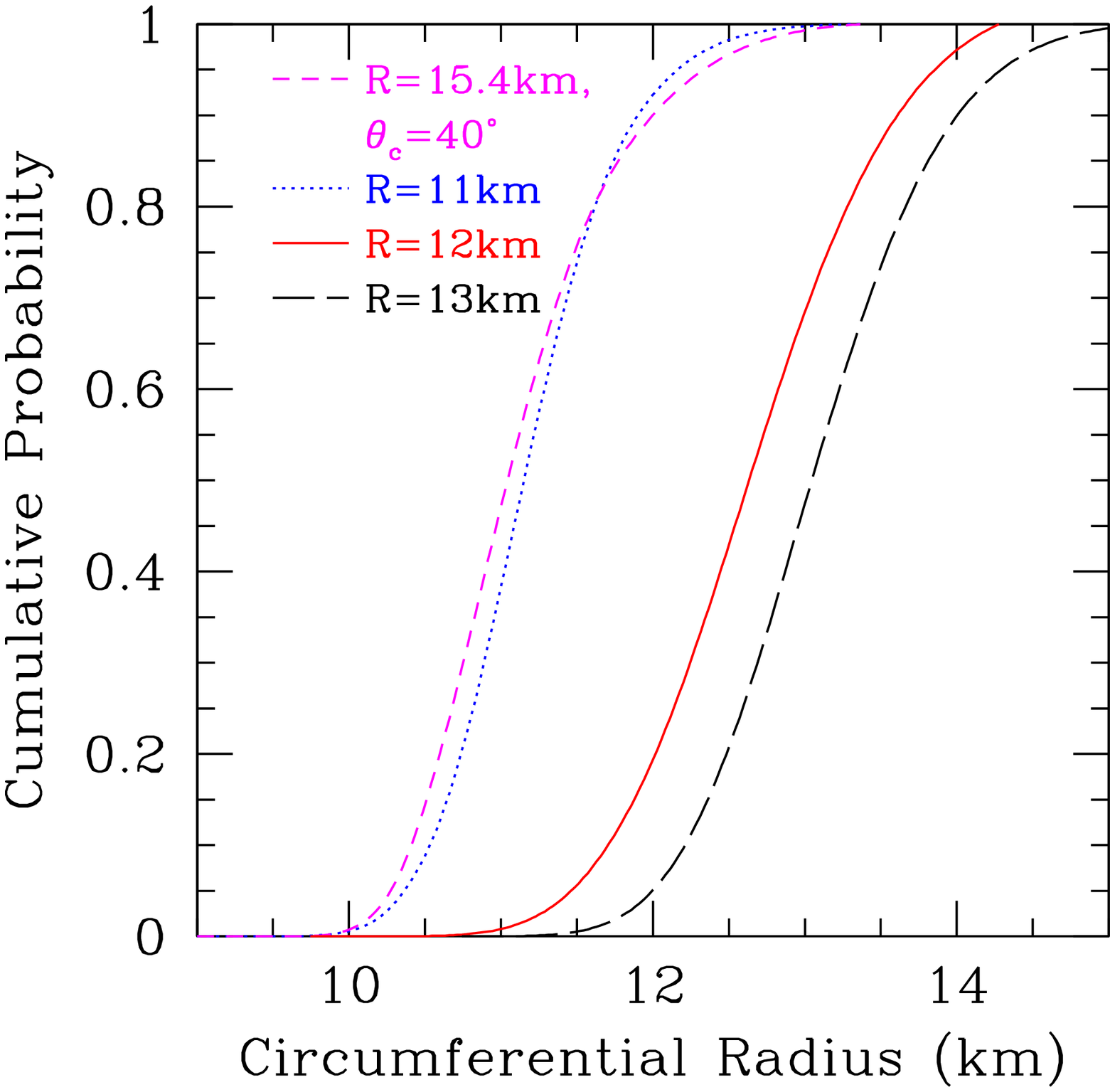}
\vskip-1.0cm
\caption{Cumulative probability of the circumferential radius for a hydrogen atmosphere model of synthetic energy-resolved {\it NICER} data for PSR~J1614$-$2230.  The lines are labeled with the radius that was assumed in the generation of the synthetic data.  See text for details of the runs.  Statistical fluctuations in the data lead to the unequal spacing of the lines.  The $R=15.4$~km, $\theta_c=40^\circ$ run was performed to test the hypothesis that it is the frms that provides the major constraints, and indeed we see that the cumulative probability distribution is almost the same as for the $R=11$~km, $\theta=90^\circ$ case, which has the same frms but different harmonic content. This figure shows that the high quality of {\it NICER} data could yield useful constraints on the radius.  We also note that at smaller radii the lines rise more steeply, because at smaller radii the maximum amount of light deflection increases more rapidly with decreasing radius than at larger radii, and hence the waveform is smoothed out more efficiently at smaller radii.
}
\label{fig:NICER}
\end{center}
\end{figure}

Because our analysis in \S~\ref{sec:XMM} showed that when $\theta_c\approx 90^\circ$, $R=11$, 12, or 13~km give good fits to the {\it XMM-Newton} data, we use those three circumferential radii as representative radii when we use $\theta_c=90^\circ$ to generate the synthetic data (note that $Rc^2/GM<3.5$ is ruled out strongly in these analyses, so that we can use $\theta_c=90^\circ$ rather than $\theta_c=85^\circ$ as we did earlier).  That is, the parameters we use to produce the synthetic data are consistent with what we know from the {\it XMM-Newton} data.  As before, we fold the data through an interstellar absorbing column of $N_H=2.4\times 10^{21}~{\rm cm}^{-2}$ (we find that marginalizing over $N_H$ gives essentially the same results).  We display the results in Figure~\ref{fig:NICER}.  We see that the likely constraints from {\it NICER} data are promising; even if $R=11$~km, a lower limit in excess of 10~km could be obtained with high significance.  The reason for the sharp lower limit is that at a fixed gravitational mass, light deflection increases strongly towards smaller radii, and thus the frms of the waveform is much less for a small radius than it is for a large radius.  We also see that the lines are not equally spaced.  This is because of statistical fluctuations in these particular realizations; we find that those fluctuations can move the best-fit radius by $\sim 0.5$~km at a true radius of 11~km, and by $\sim 0.7$~km at a true radius of 13~km.  

To test the hypothesis that it is the frms which determines the strength of the constraints, we also generated a synthetic data set with $\theta_c=40^\circ$ and $R=15.4$~km, which gives the same frms but a weaker overtone content compared with the $\theta_c=90^\circ$, $R=11$~km data set (we thank the referee for suggesting this test).  As before, we analyze the data assuming that $\theta_c=90^\circ$.  The cumulative radius distribution for this case in Figure~\ref{fig:NICER} shows that, indeed, the constraints are similar to the constraints from when we generated synthetic data using $\theta_c=90^\circ$ and $R=11$~km.  We reiterate that our assumptions produce the smallest possible radii in our analysis.  If $\theta_c<90^\circ$, or if the spot is not close to pointlike, or if there is additional unmodulated emission from the system, then small radii would be ruled out with even greater confidence.

\subsection{The effects of unmodeled modulated nonthermal emission and a biased estimate of the background}

A potential contribution which has not previously been examined in this context involves nonthermal emission that is modulated at the stellar rotational frequency.  This is a possibility because even if, as expected, the deposition of energy by magnetospheric return currents is deep enough that the emergent spectrum is thermal (see \citealt{2013ApJ...762...96B} and references therein), there are other sources of nonthermal photons that are modulated.  Indeed, the modulated gamma-rays seen using {\it Fermi} are nonthermal.  {\it NICER} data at energies $>2$~keV could be important to detect the presence of such nonthermal radiation and it is plausible that separate fits to those high energies will allow us to model out this component.  But suppose that such separate modeling is not possible.  Are there possible systematic errors in the inference of lower radius limits, which are not signaled by poor fits when we assume that the modulated component is purely thermal?

To explore this, we take our $R=13$~km simulation and add to it a modulated power-law component with a photon index $dN/dE\propto E^{-1.25}$ (this index is in the midrange of the power law fits from \cite{2012A&A...544A.108P}, although the uncertainty in the index is large).  We give this component a sinusoidal modulation that reaches zero intensity only at phase 0.5.  This contrasts with the thermal component, which has zero intensity at the middle four phases (i.e., a total of 0.25 cycles) for this radius.  Thus, the effect of the added component is to make the star seem smaller than it is. 

\begin{figure}[!htb]
\begin{center}
\plotone{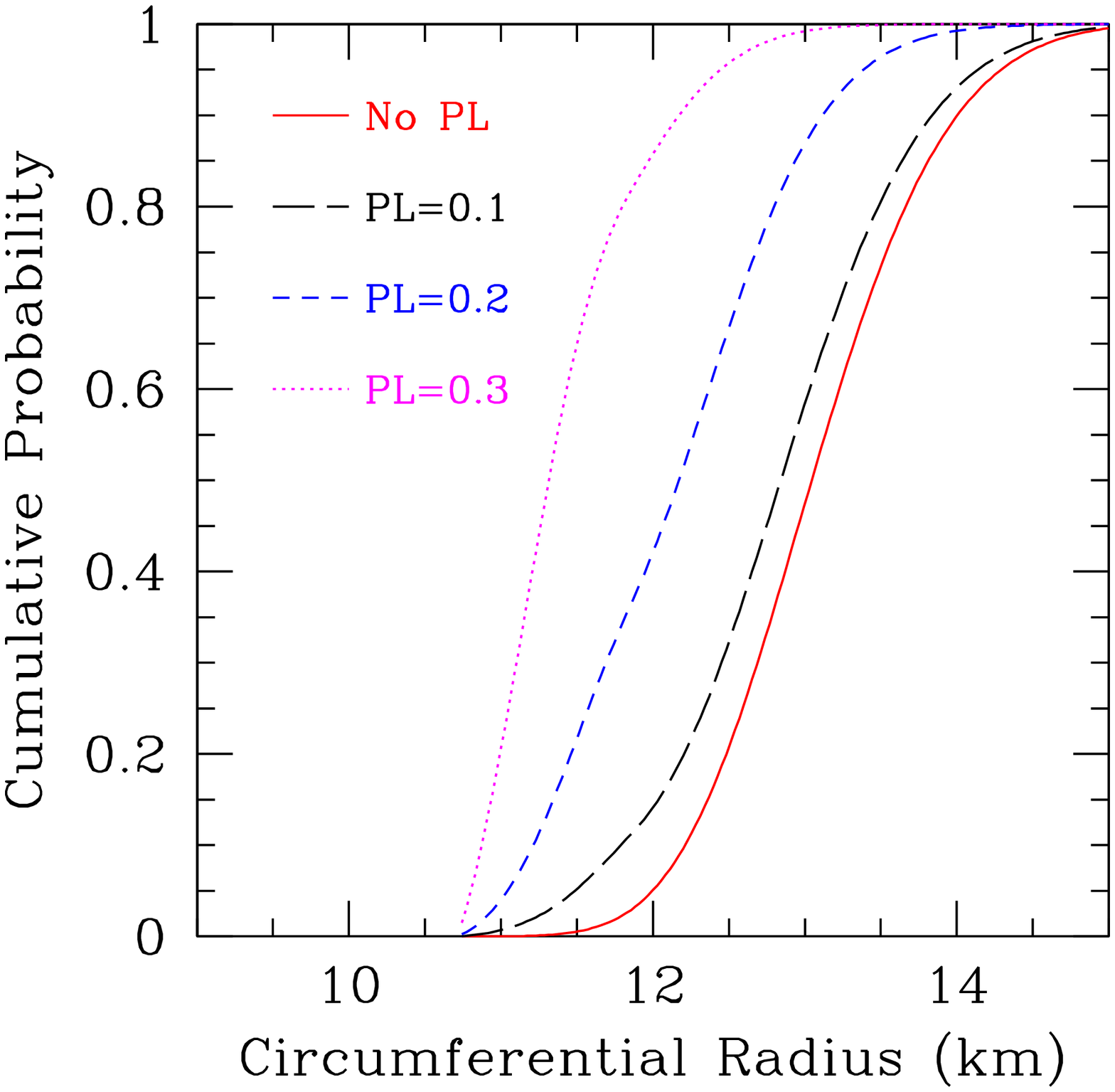}
\vskip-1.0cm
\caption{Cumulative probability distributions for the same $R=13$~km data as before, except that we have added a sinusoidal component with an absorbed $dN/dE\propto E^{-1.25}$ power law, modulated at the stellar rotational frequency, with a total number of counts in the 0.4--3~keV range that is the indicated fraction of the thermal counts from the star in this energy range.  The solid red line shows the previous no-power-law fit.  We fit the synthetic data using our usual model to determine the potential systematic errors introduced by the unmodeled modulated nonthermal component of the spectrum.  Because the thermal component is eclipsed by the star whereas the nonthermal component is an uneclipsed sinusoid, the tendency is to drive the fits towards smaller radii.  However, the unmodeled component worsens the fit: for example, the best fit for a power law with 0.3 times the thermal counts has $\chi^2=457.4$ for 387 degrees of freedom, which has a probability less than 1\%.  Thus although there would be a bias towards small radii, there would also be an indication that something is missing in the fit.  More work needs to be done, but this figure is an encouraging indicator that even a modulated but unmodeled component might not bias the inferred radius without being detectable by a poor fit.
}
\label{fig:unmodeled}
\end{center}
\end{figure}

We show the results of this calculation in Figure~\ref{fig:unmodeled}, which plots the cumulative probability distribution for the radius for three different values of the ratio of the number of modulated power-law counts in the $0.4-3$~keV range to the number of modulated thermal counts in the same energy range.  As expected, there is an increasing bias toward more compact stars for larger ratios.  If the modulation pattern of the nonthermal component were the same as the pattern of the thermal component, we would expect the bias to be smaller or even absent.  Even in the current analysis, the bias is not large in a statistical sense and in this particular example, when the modulated nonthermal component contains 30\% of the counts of the thermal component, the fit of our standard model (which does not include a nonthermal component) is poor and would therefore serve as a warning that the model is inadequate.  Thus, it could be that this is another case in which a systematic error does not lead to significant undetected biases in the waveform fitting method.

Another potential bias could be introduced if our estimate of the non-source background is systematically incorrect.  For example, if the true background at the source position has a larger number of counts than we estimate, then the true radius will be larger than we estimate because our underestimate of the background will drive our fits toward waveforms with incorrectly small modulation fractions.  Similarly, if the true background has a smaller number of counts than we estimate, then the true radius will be smaller than we find in our analysis.

\begin{figure}[!htb]
\begin{center}
\plotone{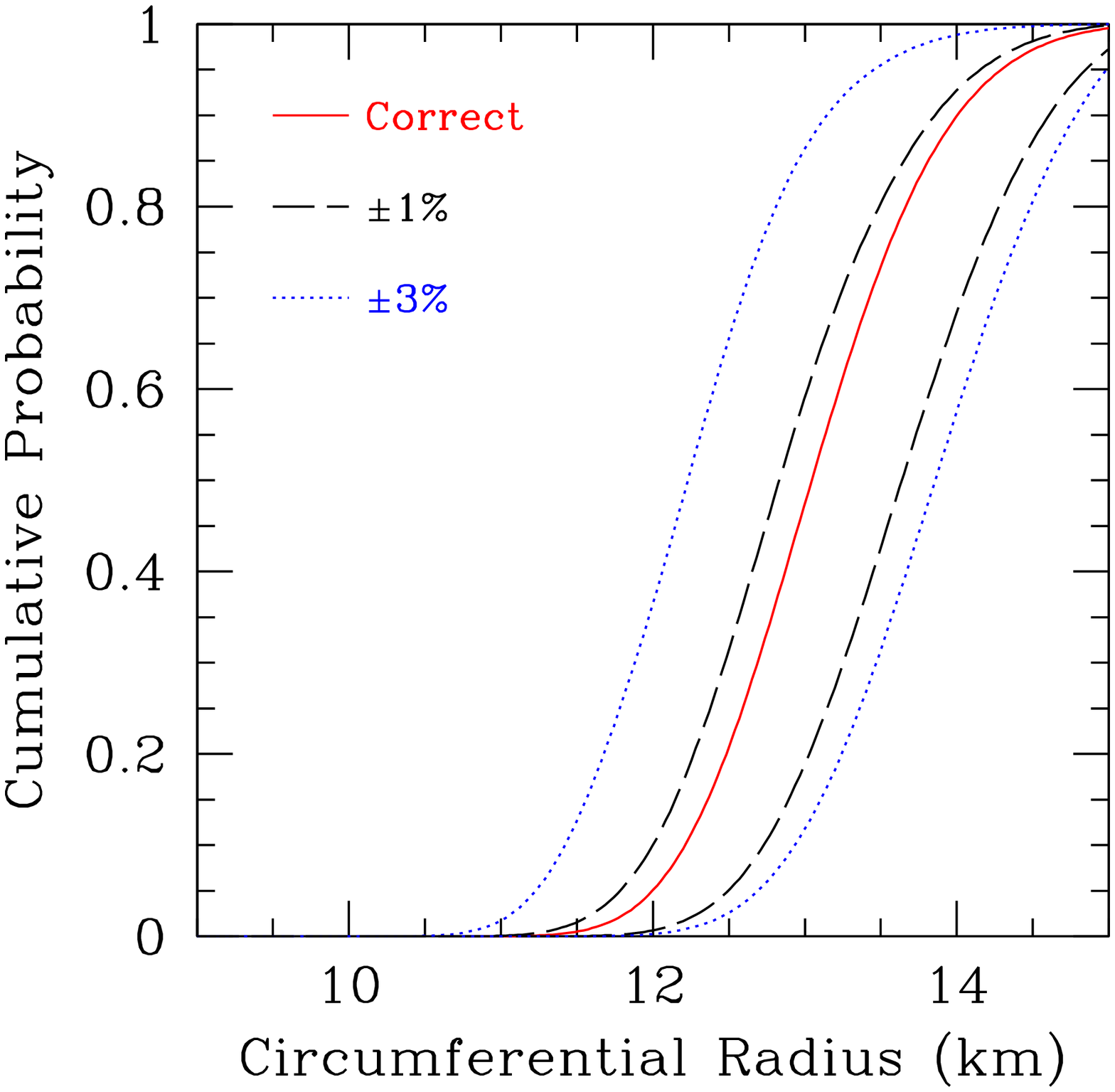}
\vskip-1.0cm
\caption{Cumulative probability distributions for our fiducial synthetic $R=13$~km data, but using in the analysis a model for the synthetic background that differs from the correct background by a factor that is the same at all energies: 0.97 (left blue dotted line), 0.99 (left black dashed line), 1.01 (right black dashed line), or 1.03 (right blue dotted line). The solid red line shows the previous fit, which uses the best estimate of the background derived from 0.1 Ms of synthesized off-source data.  Deviations of more than $\sim$3\% from the true background produce poor fits that would be detectable in the analysis; for example, a background that is 10\% too small (large) gives a $\chi^2$ that is 90 (143) larger than the $\chi^2$ for the true background.  This figure shows that systematic errors introduced by incorrect background estimates are modest (a few tenths of a kilometer) if the fit is good.  Note that for a given fractional deviation from the true background level an incorrectly high background is easier to detect than an incorrectly low background, because for a given frms in the data, added background strongly worsens the fit whereas a deficit in the background can be accommodated by making the star smaller.  This bias, if it exists, is therefore more likely to lead to an underestimate than to an overestimate of the radius.
}
\label{fig:back}
\end{center}
\end{figure}

To explore this, we analyze our $R=13$~km data using background estimates that are multiplied by some factor compared to the best guess.  For our assumed number of counts, background estimates that are more than $\sim 3$\% different from the true background lead to poor fits, so we investigate the effects of changing the background by $\pm 1$\% and $\pm 3$\%, equally at all photon energies.  We show the results in Figure~\ref{fig:back}.  Again, the bias for well-fit data is at most a few tenths of a kilometer.  We also note that, for a given fractional deviation from the true background, an increase in the background produces a significantly worse fit than a decrease in the background.  This is because the observed waveform has a given amount of modulation, so the addition of unmodulated background produces a less-modulated waveform and thus a poorer fit.  Therefore, this is a bias that is more likely to lead to an underestimate than an overestimate of the radius.

\section{DISCUSSION AND CONCLUSIONS}
\label{sec:summary}

{\it NICER} observations will have unprecedented soft X-ray timing precision, which is well-matched to the properties of non-accreting neutron stars.  Thus, these observations will produce strong constraints on the masses and radii of those stars, and hence on the properties of cold matter beyond nuclear density.  We argue here that data from PSR~J1614$-$2230 could play an important role in those constraints, because although upper limits to the radius will be difficult given the low count rate, lower limits to the radius could be quite constraining.  These limits require good understanding of the non-source background (in contrast to the limits that will be obtained from higher count rate sources such as PSR~J0437$-$4715, which will not require precise knowledge of the background).  Thus to support these observations, it would be useful to have {\it Chandra} observations of blank fields near the source to assess the degree of variation from field to field and thus the degree to which we could be confident about the non-source background during on-source observations.  

Our simulations have assumed a time of 0.5~Ms on-source combined with 0.1~Ms used to determine the non-source background.  An initial short on-source observation (say, 50~ks) would be useful to verify that there are in fact significant pulsations at the rotational frequency.  More time on background would be inefficient for a given total time devoted to this source.  Less time on background would lead to greater statistical uncertainties, but even more importantly would lead to questions about the uniformity of the background; for this reason, we recommend that the 0.1~Ms of background observations be divided into pointings over several nonoverlapping fields that are near the source but do not include it.  There is an existing {\it Chandra} observation of the source from 2007 (ObsID 7509), and the $\sim$90~ks of {\it XMM-Newton} MOS images of the surrounding field give some information about the uniformity of the background (thanks to S. Bogdanov for bringing this to our attention), but to achieve the desired precision it would be very helpful to have new {\it Chandra} observations as well as {\it NICER} observations of the surrounding fields.  More time devoted to PSR~J1614$-$2230 would of course improve the precision of the constraints; for example, \citet{2013ApJ...776...19L} find that for a given emission geometry and total number of counts $N_{\rm tot}$ the precision scales as ${\cal R}^{-1}$, where ${\cal R}\equiv N_{\rm osc}/\sqrt{N_{\rm tot}}=1.4~{\rm frms}\sqrt{N_{\rm tot}}$ and $N_{\rm osc}$ is the total number of oscillating counts.

In summary, we find that if we know the background, then {\it NICER} observations of PSR~J1614$-$2230 could produce conservative lower limits on the radius with statistical uncertainties that range from $\sim 0.5$~km at low radii to $\sim 0.7$~km at high radii.  If those lower limits exceed 11~km, then the observations will contribute significantly to our understanding of the matter in neutron stars.

\acknowledgements

This work was supported in part by NASA NICER grant NNX16AD90G.  We thank Zaven Arzoumanian, Slavko Bogdanov, Deepto Chakrabarty, Fred Lamb, Scott Lawrence, Denis Leahy, Sharon Morsink, Simin Mahmoodifar, Joonas N\"attil\"a, Juri Poutanen, Paul Ray, and Natalie Webb for useful discussions regarding this work, and Natalie Webb for kindly providing the {\it XMM-Newton} pn data on PSR~J1614$-$2230.  We also thank the referee for a helpful and timely report.

\bibliography{ms}

\end{document}